\documentclass[twocolumn,showpacs,preprintnumbers,amsmath,amssymb]{revtex4}

\usepackage{graphicx}
\usepackage{dcolumn}
\usepackage{bm}

\begin{document}

\title{Large-scale dynamical simulations of the three-dimensional  XY spin glass}
\author{ Qing-Hu Chen$^{1,2}$}
\address{
$^{1}$ Center for Statistical and Theoretical Condensed Matter
Physics, Zhejiang Normal University, Jinhua 321004, P. R. China \\
$^{2}$ Department of Physics, Zhejiang University, Hangzhou 310027,
P. R. China}

\date{\today}

\begin{abstract}
Large-scale simulations have been performed in the current-driven
three-dimensional XY spin glass  with resistively-shunted junction
dynamics for sample sizes up to $64^3$. It is observed that the
linear resistivity at low temperatures tends to zero, providing a
strong evidence of a finite temperature phase-coherence (i.e.
spin-glass) transition. Dynamical scaling analysis demonstrates that
a  perfect collapse of current-voltage data can be achieved. The
obtained critical exponents agree  with those in equilibrium Monte
Carlo simulations, and are compatible with those observed in various
experiments on high-T$_c$ cuprate superconductors. It is suggested
that the spin and the chirality order simultaneously. A genuine
continuous depinning transition is found at zero temperature. For
low temperature creep motion, critical exponents are evaluated, and
a non-Arrhenius creep motion is observed in the low temperature
ordered phase.  It is proposed that the XY spin glass gives an
effective description of the transport properties in high-T$_c$
superconductors with d-wave symmetry.

\end{abstract}

\maketitle

\section{Introduction}

The ordering of the three-dimensional (3D) XY spin glass model
\cite{Jain} has attracted considerable attentions.  Theoretically,
earlier work suggest transitions at  zero or very low temperatures.
Following the pioneering work by Villain\cite{Villain} where the
role of chirality (i.e. vorticity) arising for non-collinear XY
system, was emphasized, Kawamura et al proposed a chiral-glass
transition, even though the finite temperature spin-glass transition
vanishes\cite{Kawamura0}.  Later on, by calculating the Binder ratio
and the spin-overlap distribution function for lattice sizes up to
$L=16$, Kawamura and Li have given a numerical evidence that the low
temperature phase is a chiral glass  without the conventional spin
glass order\cite{Kawamura1}.

However, this  spin-chirality decoupling scenario is not consistent
with  several
studies\cite{Maucourt,Matsubara,Akino,Lee,Granato1,Granato2}.
Maucourt and Grempel\cite{Maucourt} and subsequently, Akino and
Kosterlitz\cite{Akino} found evidence for finite temperature
spin-glass transition from zero-temperature domain wall
calculations.  By performing a finite-size scaling analysis of the
correlation length $\xi$, the most successful technique developed in
the Ising spin glass\cite{Ballesteros},   Lee and Young \cite{Lee}
observed a transition at the same temperature for both spins and
chiralities for lattice sizes up to $L=12$. Beside the equilibrium
simulation, the resistive behavior in the XY spin glass has been
also studied. In a vortex representation, Wengle and Young showed
evidence of a resistive transition at finite
temperatures\cite{Wengel} for lattice sizes up to $L=10$. Granato
observed that the current-voltage characteristic in the XY spin
glass with both bimodal  and Gaussian couplings   for the lattice
size $L=12$ exhibited scaling behaviour\cite{Granato1,Granato2},
which was interpreted  in terms of both the phase-coherence
(spin-glass) transition  and the chiral-glass transition.

The controversy on the spin-chirality decoupling scenario is however
still ongoing\cite{Youngcm}. For the Heisenberg spin glass, the
transitions at the same temperature was also observed for both spins
and chiralities in Ref. \cite{Lee} with the use of the same method
for XY spin glass. However, subsequently, it is found that the
situation becomes rather unclear when results of larger sizes are
included\cite{Hukushima,Campos,Lee1,Viet,Fernandez}. The data at the
low temperatures in larger samples show rather marginal behavior,
i.e. the system is close to the lower critical dimension. Recently,
there seems a intensely competition for the lattice sizes accessible
\cite{Lee1,Viet,Fernandez}, the record until now is
$L=48$\cite{Fernandez}. Motivated from the debate for Heisengurg
case, Young and his collaborator further study the XY spin glass
with  larger lattice sizes up to $L=24$\cite{Pixley}, and observed
similar marginal behavior for low temperatures and large systems. So
it is also desirable to perform the simulations on very large sample
for XY spin glass. In contrast with the Ising and Heisenberg spin
glass, there is an additional technique in XY spin glass, that is
the measurement  of the resistivity from dynamical
simulations\cite{Wengel,Granato1,Granato2}. Moreover, one can access
larger system in this kind of dynamical simulations, because a
steady state is more easily reached than equilibration duo to the
presence of the external driven force.  It is also of interest to
know whether the previous picture based on small systems ($L\le 12$)
in the dynamical simulation \cite{Wengel,Granato1,Granato2} is
modified on larger systems for the XY case.

On the one hand, the XY spin glass has found experimental
realization, not only in layered  manganite, e.g.
$Eu_{0.5}Sr_{1.5}MnO_{4}$ \cite{Mathieu}, but also in high-T$_c$
cuprate superconductors with the d-wave paring symmetry
\cite{Sigrist}, which can be regarded a random distribution of $\pi$
junctions\cite{Kawamura3,Matsuura,Yamao,Papadopoulou,Li}. The nature
of  d-wave  symmetry will changes the sign of the coupling between
XY spins, while the spin angle denotes the phase of the
superconducting order parameters.  So the XY spin glass  is expected
to be used to interpret the phenomena observed in high-T$_c$ cuprate
superconductors, such as vortex glass
phase\cite{theory,FFH,Nattermann},  a true superconducting state
with vanishing linear resistivity. The evidences to support the
existence of this phase have been reported in many
experiments\cite{exp,exp1}.

Since the  chiral variable in the XY spin glass can  be defined as
the vortex in the plaquette\cite{Granato1,Granato2}, some techniques
developed in superconducting vortex models\cite{luo, chen1,chen2}
can be in turn  employed to characterize  the nature of the low
temperature ordered state  in the XY spin glass. As is well known,
in the random pinning environment, the energy landscape for the
vortex motion is highly nontrivial. The theoretical understanding
for the nonlinear dynamics response has been advanced in many
years\cite{LO}. But the full theoretical study is still very
challenging, computer simulations are hopeful to provide useful
insights.

In this paper, by resistively-shunted-junction dynamics, we perform
large-scale dynamical simulations in the 3D spin glass, both the
phase-coherence transition temperature $T_g$ and the critical
exponents are estimated. The depinning transition at
zero-temperature and creep motion  below  $T_g$ are also
investigated. The rest of the paper is organized as follows. Sec.II
describes the model and dynamic method. Sec.III presents our main
results, where some discussions are also performed. Finally, a short
summary is given in the last section.

\section{Model and dynamic method}

The Hamiltonian  of the 3D XY spin glass in the phase representation
is given by \cite{Lee}
\begin{equation}
H=-\sum_{\langle ij\rangle }J_{ij}\cos (\phi _{i}-\phi _{j}).
\label{Hamil}
\end{equation}
where the sum is over all nearest neighbor pairs on a 3D square
lattice, $\phi _{i}$ specifies the phase of the superconducting
order parameter on grain $i$, $J_{ij}$ denotes the strength of
Josephson coupling between neighboring grains with zero mean and
standard deviation unit. The present simulations are performed with
the lattice size $L=64$ for all directions, considerably larger than
those in literature.

The Resistivity-Shunted-Junction dynamics is incorporated in
simulations, which can be described as
\begin{equation}
{\frac{\sigma \hbar }{2e}}\sum_{j}(\dot{\phi _{i}}-\dot{\phi _{j}})=-{\frac{%
\partial H}{\partial \phi _{i}}}+J_{{\rm ext},i}-\sum_{j}\eta _{ij},
\end{equation}
where $J_{{\rm ext},i}$ is the external current which vanishes
except for the boundary sites. The $\eta _{ij}$ is the thermal noise
current with zero mean and a correlator $\langle \eta _{ij}(t)\eta
_{ij}(t^{\prime })\rangle =2\sigma k_{B}T\delta (t-t^{\prime })$. In
the following, the units are taken of $2e=\hbar =\sigma =k_{B}=1$.

In the present simulations, a uniform external current $I_{x}$ along
$x$-direction is fed into the system, the fluctuating twist boundary
condition \cite{chen1} is applied in the $xy$ plane, and the
periodic boundary condition is used in the $z$ axis. In the $xy$
plane, the supercurrent between sites $i$ and $j$ is now given by $\
J_{i\rightarrow j}^{(s)}=J_{ij}\sin (\theta _{i}-\theta
_{j}-A_{ij}-{\bf r} _{ij}\cdot {\bf \Delta })$, with ${\bf \Delta
}=(\Delta _{x},\Delta _{y})$ the fluctuating twist variable and
$\theta _{i}=\phi _{i}+{\bf r}_{i}\cdot {\bf \Delta }$. The new
phase angle $\theta _{i}$ is periodic in both $x$- and
$y$-directions. Dynamics of ${\bf \Delta }_{\alpha }$ can be then
written as
\begin{equation}
\dot{\Delta}_{\alpha }={\frac{1}{L^{3}}}\sum_{<ij>_{\alpha
}}[J_{i\rightarrow j}^{(s)}+\eta _{ij}]-I_{\alpha }, \alpha =x,y
\label{delta-dot}
\end{equation}
The voltage drop is $V=-L{\dot{\Delta}_{x}}$.

The above equations can be solved efficiently by a pseudo-spectral
algorithm \cite{chen2} due to the periodicity of phase in all
directions. The time stepping is done using a second-order
Runge-Kutta scheme with $\Delta t=0.05$. Our runs are typically
$(4-8)\times 10^{7}$ time steps and the latter half time steps are
for the measurements.  Our results are based on one realization of
disorder. The  present system size is much too larger than those
reported in literature,  a good self-averaging effect is expected.
We have performed an additional simulations with a different
realization of disorder for further confirmations, and observed
quantitatively the same behavior. Note that it  is practically hard
to perform any serious disorder averaging for the present very large
system size.

\section{Simulation results and discussions}

\subsection{Finite temperature phase-coherence transition}

\begin{figure}[tbp]
\centering
\includegraphics[width=7cm]{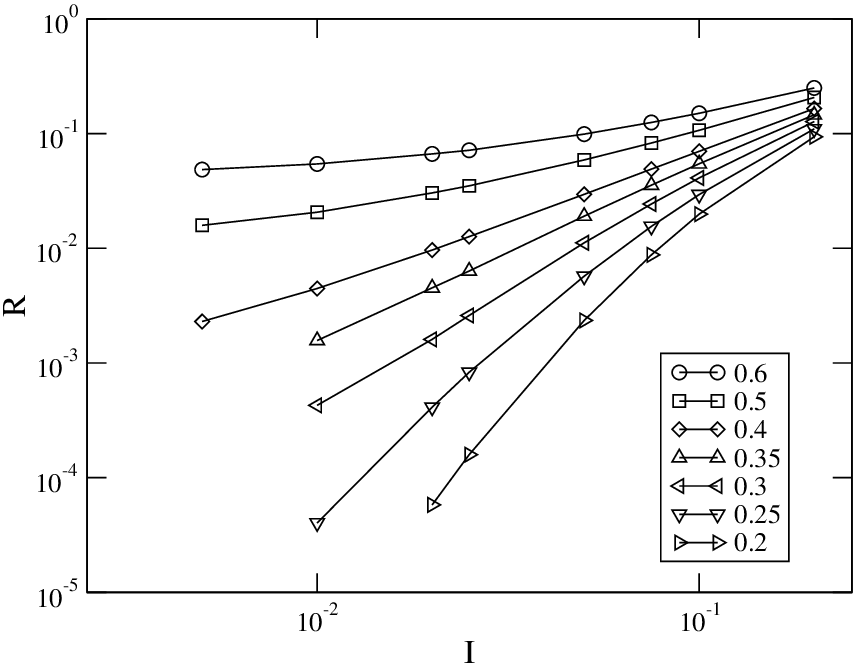}
\caption{Current-resistivity curves  at various temperatures}
\label{Fig1}
\end{figure}

First, we study the phase-coherence  transition. The current-voltage
characteristics are simulated at various temperatures ranged from
[0.2, 0.6]. At each temperature, we try to probe the system at a
current as low as possible. The voltage is determined when a steady
state is arrived. Fig. 1 demonstrates the resistivity $R=V/I$ as a
function of current $I$ at various temperatures. It is clear that,
at lower temperatures, $R$ tends to zero as the current decreases,
which follows that there is a true superconducting phase with zero
linear resistivity. While $R$ tends to a finite value at higher
temperatures, corresponding to an Ohmic resistivity. These
observations provide a strong evidence of occurrence of a
phase-coherence transition at finite temperature in the 3D XY spin
glass in the dynamical sense.

For a continuous phase transition characterized by the divergence of
the characteristic length and time scales $t\sim\xi^{z}$ (z is the
dynamic exponent) , Fisher, Fisher, and Huse \cite {FFH} proposed
the following dynamic scaling ansatz
\begin{equation}
TR\xi^{z+2-d}=\Psi_{\pm }(I\xi^{d-1}/T). \label{ffh}
\end{equation}
where $d$ is the dimension of the system ($d=3$ in this paper),
and $\xi\propto\mid T/T_g-1\mid^{-\nu}$  is the correlation length
which diverges at the transition. $\Psi(x)$ is a scaling function,
with + and - signs corresponding to $T>T_g$ and $T<T_g$. Eq.
(\ref{ffh}) was often used to scale measured current-voltage data
experimentally\cite{exp,exp1}.

\begin{figure}[tbp]
\centering
\includegraphics[width=7cm]{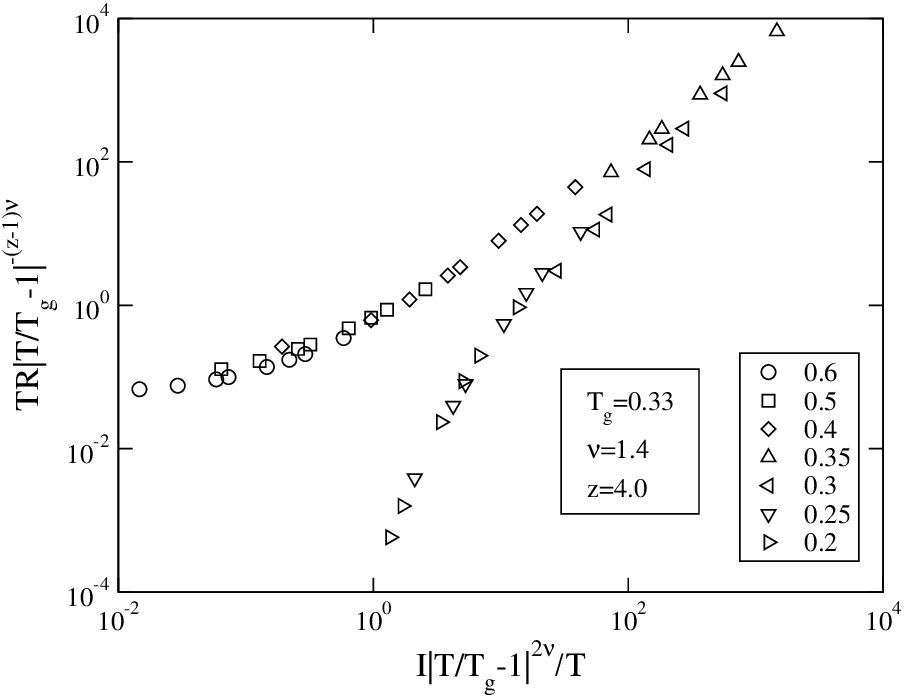}
\caption{Dynamic scaling of current-voltage data  at various
temperatures according to Eq. (\ref{ffh}). } \label{Fig2}
\end{figure}

To extract the critical behavior from the numerical results of the
current-voltage characteristics, we will also perform a dynamical
scaling analysis.  As shown in Fig. 2, using $T_{g}=0.33\pm 0.02$,
$z=4.0\pm 0.1$, and $\nu =1.4\pm 0.1$, an excellent collapse is
achieved  according to  Eq. (\ref{ffh}). The value of $\nu$ is by no
means close to the Ising value $2.15\pm 0.15$\cite{Ballesteros},
demonstrating the XY and Ising spin glass belong to different
universality classes.

The finite-size effect is particularly significant at temperatures
sufficiently close to $T_{g}$ when the correlation length exceeds
the system size. For the temperatures considered  and the very
lattice size $L=64$ here, we believe that the finite-size effect is
negligible in the present simulations. To confirm this point, we
perform particular simulations right at $T_{g}=0.33$ obtained above
for different system size. At $T_g$, the correlation length is cut
off by the system size in any finite system, the scaling form
(\ref{ffh}) for $d=3$ becomes
\begin{equation}
T_{g}RL^{z-1}=\Psi (IL^2/T_{g}). \label{ffhtg}
\end{equation}
As shown in  Fig. 3, a good collapse is shown using $z=4.2 \pm 0.1$.
This consistence demonstrates that the estimate from Fig. 2 is
reliable. Therefore a new evidence of a finite-temperature
phase-coherence transition is provided convincingly in the 3D XY
spin glass.

\begin{figure}[tbp]
\centering
\includegraphics[width=7cm]{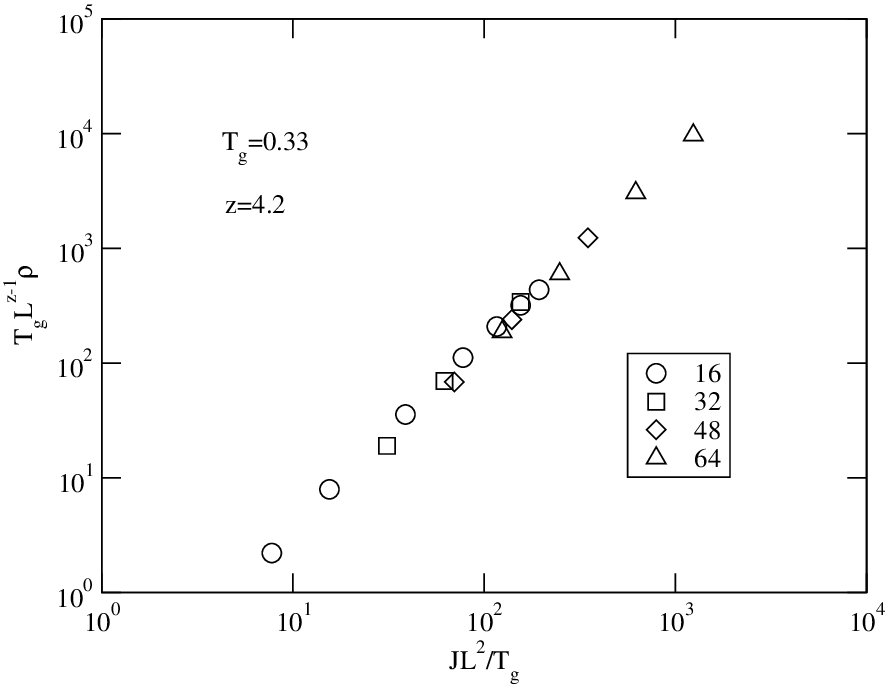}
\caption{Dynamic scaling of current-voltage data  at $T_{g}=0.2$
according to Eq. (\ref{ffhtg}).} \label{Fig3}
\end{figure}

The above obtained $T_{g}$ and static exponent $\nu $ are consistent
with  $T_g=0.34\pm0.02$ and $\nu =1.2\pm 0.2$ obtained in
equilibrium Monte Carlo simulations\cite{Lee} for lattice sizes up
to $L=12$. More recently, the data for the spin-glass and
chiral-glass correlation length  at larger sizes show marginal
behavior, indicating that the lower critical dimension is close to
$3$ for both spins and chiralities\cite{Pixley}. The crossing
temperatures for the spin and chiral-glass correlation length
decrease very slightly with the size with large uncertainties, close
to $0.3$ for largest lattice sizes they can simulate. The data for
the ratio of the chiral and spin-glass correlation for the largest
sizes intersects for $T$ about $0.33$ and then splay out in the low
temperature side. The size dependence of the crossing temperature at
large sizes is very weak, supporting a single transition for both
spins and chiralities. Interestingly, this crossing temperature is
in excellent agreement with the present $T_{g}$ for a very large
lattice size. We do not think it is a coincidence.

Previous driven Monte Carlo dynamical simulation on the same model
for a small lattice size ($L=12$) \cite{Granato1} estimated
$T_g=0.335$ , $\nu =1.2$, and $z=4.4$, basically consistent with the
present ones. It follows that the results are not sensitive to the
system size in the dynamical simulations, even the detailed dynamics
is different. The Monte Carlo dynamical simulations in the vortex
representation also show an equilibrium transitive
transition\cite{Wengel}. The estimated value for $\nu=1.3\pm0.3$
agrees  with the present one. But the dynamic exponent $z\approx
3.1$ is obvious lower, indicating different dynamic universality
class for the XY spin glass in the phase and vortex representations.

Typical values of the correlation length exponent and the dynamical
exponent extracted from the various experiments on high-T$_c$
cuprate superconductors\cite{exp,exp1} fall in the range of
$\nu\approx 1.0-1.7$ and $z\approx 4.0-5.0$. Our results for these
static and dynamic exponents in the XY spin glass are consistent
with these experimental ones. In addition, nonlinear resistivity
measurement in the $YBa_{2}Cu_{4}O_8$ bulk sample\cite{Yamao} near
the onset of the paramagnetic Meissner effect have been interpreted
as a chiral-glass transition owing to the presence of $\pi$
junctions. The nonlinear contribution $\rho_2 $ to the resistivity
was observed to have a peak at the transition with power-law
behavior $\rho\sim J^{-\alpha}$ and exponent $\alpha=1.1 \pm 0.6$,
which has been reproduced by a generalized XY spin glass
model\cite{Li}. Granato proposed that it could also be interpreted
as a consequence of the underlining phase coherence transition and
the exponent $\alpha$ is determined by the dynamic exponent through
$\alpha=(5-z)/2$\cite{Granato1}. By the above obtained dynamic
exponent, we  have  $\alpha=0.5\pm 0.05$,   compatible with the
experimental value within the  error bar. Note that the present
value of $z$ in the large-scale simulations is smaller than  those
previously  given  by Granato\cite{Granato1,Granato2} by about
$10\%$,  so our estimated $\alpha$ is more close to experimental
one. Based on  the above comparisons with experiments, we propose
that the XY spin glass gives an effective description of the
transport properties near the phase-coherence transition in d-wave
superconductors.

Until now, at least there is a general consensus in 3D XY spin glass
that the chiral-glass transition occurs at a finite temperature
larger than or equal to the spin-glass transition temperature $T_g$
\cite{Kawamura1,Maucourt,Matsubara,Akino,Lee,Granato1,Granato2,Pixley}.
The data for the chiral correlation length in the recent large-scale
equilibrium Monte Carlo simulations\cite{Pixley} reveals that the
chiral-glass transition temperature is around $0.30-0.33$, not
larger than the present  spin-glass transition temperature. We
therefore suggest that the spin and the chirality probably order
simultaneously.

\subsection{Depinning transition and creep motion}

\begin{figure}[tbp]
\centering
\includegraphics[width=7cm]{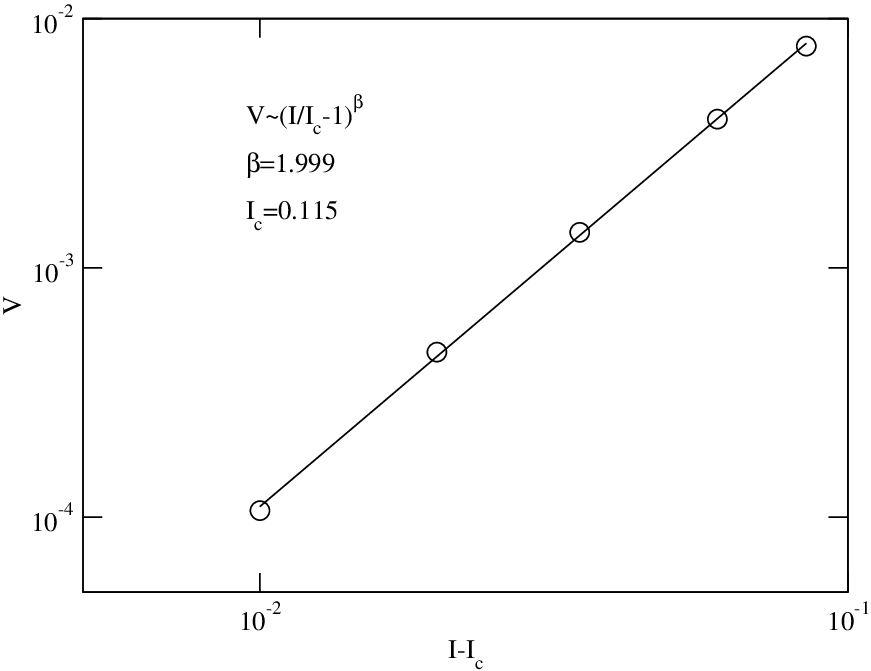}
\caption{Log-log plots of $V-I$ curve at zero-temperatures.}
\label{Fig4}
\end{figure}
With the low temperature spin-glass phase in hand, in the remaining
part of the paper, we will study the depinning and creep phenomena
in this phase.

To study the depinning transition at zero temperature, we start from
high currents with random initial phase configurations.  The current
is then lowered step by step.  Fig. \ref{Fig4} shows the
current-voltage characteristics at $T=0$. Interestingly, we observe
continuous depinning transitions with unique depinning currents,
which can be described as $V\propto (I-I_{c})^\beta$ with
$I_{c}=0.115\pm 0.001$, $\beta=1.99\pm 0.02$. Note that the
depinning exponent $\beta$ is greater than $1$, consistent with the
mean field studies on charge density wave models\cite{Fisher}.

At low temperatures,  the current-voltage characteristics is rounded
near the zero-temperature critical current due to thermal
fluctuations. An obvious crossover between the depinning and creep
motion can be observed around $I_{c}$ at the lowest accessible
temperatures. In order to address the thermal rounding of the
depinning transition, Fisher\cite{Fisher} first suggested to map
this system to the ferromagnet in fields where the second-order
phase transitions occur. This mapping  was latter extended to the
random-field Ising model\cite{Roters}, flux lines in type-II
superconductors\cite{luo}, and Josephson junction arrays\cite{liu}.
If the voltage is identified as the order parameter, the current and
temperature are identified as the inverse temperature and the field
in the ferromagnetic system respectively, analogous to the
second-order phase transitions, a scaling relation among the
voltage, current, and temperature reads
\begin{equation}~\label{scaling1}
V(T,I)=T^{1/\delta}S[T^{-1/\beta\delta}(1-I_{c}/I)],
\end{equation}
where $S(x)$ is a scaling function.

\begin{figure}[tbp]
\centering
\includegraphics[width=7cm]{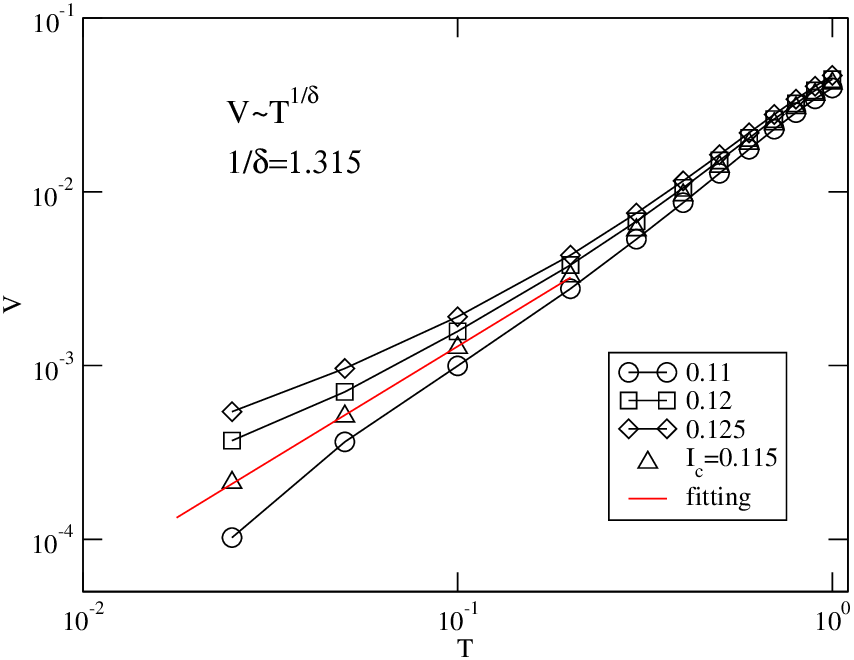}
\caption{Log-log plots of $V-T$ curves at three currents around
$I_{c}$.} \label{fig5}
\end{figure}

Eq. (6) reveals that right at $I_{c}$ the voltage shows a power-law
behavior $V(T,I=I_{c})\propto T^{1/\delta}$, providing a method to
determine the critical exponent $1/\delta$. The  $V-T$ curves at
three currents are presented in Fig. \ref{fig5} on a  log-log plot.
The critical current is seen to be between $0.11$ and $0.125$. The
values of voltage at other currents within $(0.11,0.125 )$ can be
evaluated by quadratic interpolation. The square deviations from the
power law can be calculated. The critical current is defined at
which the square deviation is minimum. We obtain $I_{c}=0.115\pm
0.02$, consistent with those obtained above at zero temperature. The
temperature dependence of voltage at the critical current is also
plotted in Fig. \ref{fig5}, yielding $1/\delta=1.315\pm 0.001$.

\begin{figure}[tbp]
\centering
\includegraphics[width=7cm]{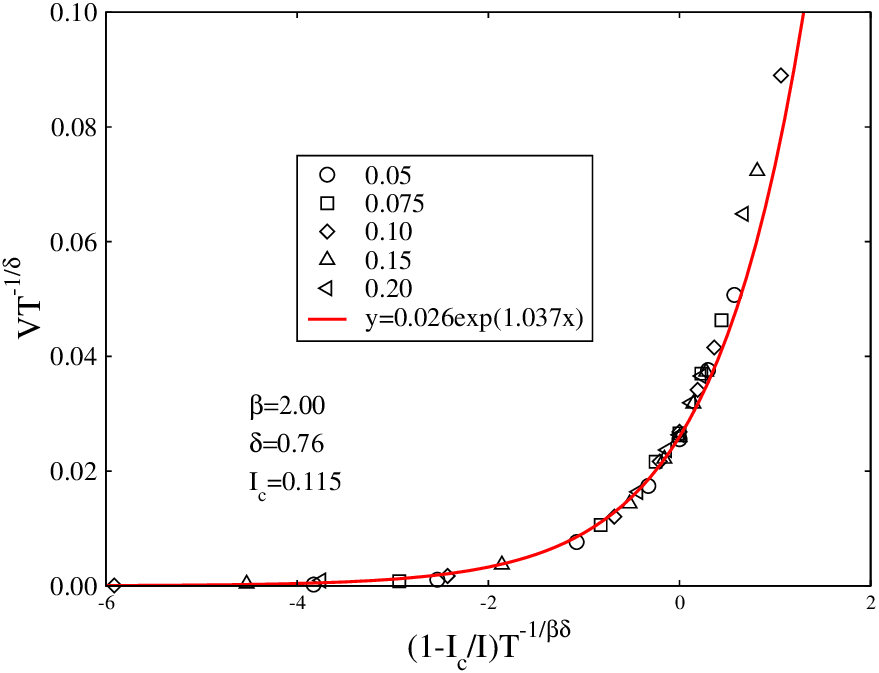}
\caption{Scaling plot  of the current-voltage data at  various
temperatures below $T_g$ according to Eq. (\ref{scaling1}).}
\label{fig6}
\end{figure}

With the critical exponent $\delta$ and the critical current
$I_{c}$, we can adjust the depinning exponent $\beta$ to achieve the
best data collapse according to the scaling relation Eq.
(\ref{scaling1}) for $I\le I_c$. In Fig. \ref{fig6}, a optimal data
collapse of the current-voltage  data at various temperatures below
$T_g$ provides an estimate of  $\beta=2.00 \pm 0.01$, which is in
excellent agreement with those derived at $T=0$ depinning
transition. Moreover, the scaling function with the form $V \propto
T^{1/\delta }\exp [A (1-\frac{I_c}I)/T^{\beta\delta}]$ is used to
fit well the current-voltage data in the creep regime, which is also
list in the legends of Fig. \ref{fig6}. As we know, the product of
the two exponents $\beta\delta$ describes the temperature dependence
of the creeping law. $\beta\delta \approx 1.52$ deviates from unity,
demonstrating that the creep law is a non-Arrhenius type. The
non-Arrhenius type creep behaviors have been also previously
observed in charge density waves\cite{Middleton}.

In the  3D XY spin glass, the existence of a stable spin glass phase
at finite temperature is well established through both previous
equilibrium studies\cite{Lee,Pixley} and present dynamical
simulations. The value of $\beta\delta$ is close to $3/2$, similar
to the 3D vortex glass and gauge glass model\cite{chen1}. We believe
that the value of $\beta\delta$ is generally  $3/2$ in the glass
phases in  various strongly disordered 3D frustrated XY model. The
further analytical work is highly called for.

\section{conclusions}

We have performed large scale dynamical simulations on the 3D XY
spin glass  within the  resistively-shunted-junction dynamics.  The
strong evidence for the low temperature spin-glass phase is provided
in dynamical sense. By the dynamical scaling analysis, a perfect
collapse of simulated current-voltage data is achieved by using
$T_{g}=0.33\pm 0.02$ , $z=4.0\pm 0.1$, and $\nu =1.4\pm 0.1$. These
critical values agree with those in the previous equilibrium Monte
Carlo simulations as well as  driven Monte Carlo dynamical
simulations. In the resistive simulations, the size dependence of
various exponents and critical temperature in the phase-coherence
transition are very weak at larger size, providing a good technique
to measure critical properties in this model. The static and dynamic
exponents are compatible   with those observed in the various
experiments on high-T$_c$ cuprate superconductors, which underlines
the significance of XY spin glass in  d-wave superconductors.
Combined with the recent equilibrium Monte Carlo simulations, we
suggest a single glass transition involving both spins and
chiralities.

We also study the depinning transition at zero temperature and creep
motion at low temperatures in detail. A genuine continuous depinning
transition is observed and the depinning exponent is evaluated. With
the notion of scaling, the critical exponents are estimated, which
are consistent with those from independent simulations at zero
temperature and at critical current. The value of $\beta\delta$ is
close to $3/2$ and the scaling curve is fitted well by a exponential
function, suggesting a non-Arrhenius type creep motion in the low
temperature ordered phase in  the   XY spin glass.

Finally, it is proposed that the XY spin glass model may capture the
essential transport feature in high-T$_c$ cuprate superconductors
with  d-wave symmetry. Further experimental and theoretical works
are clearly motivated.

\section{Acknowledgements}

This work was supported by National Natural Science Foundation of
China under Grant No. 10774128, PCSIRT (Grant No. IRT0754) in
University in China,  National Basic Research Program of China
(Grant Nos. 2006CB601003 and 2009CB929104), Zhejiang Provincial
Natural Science Foundation under Grant No. Z7080203, and Program for
Innovative Research  Team  in Zhejiang Normal University.

\end{document}